\documentclass[11pt]{article}

\usepackage[T1]{fontenc}
\usepackage[utf8]{inputenc}

\usepackage[
    a4paper,
    margin=2.5cm
]{geometry}

\usepackage{amsmath}
\usepackage{amsfonts}
\usepackage{amssymb}

\usepackage{graphicx}
\usepackage{booktabs}
\usepackage{multirow}
\usepackage{tabularx}

\usepackage{cite}

\usepackage{tikz}
\usetikzlibrary{
    arrows.meta,
    positioning,
    fit,
    shapes.geometric
}

\usepackage{microtype}

\usepackage[hidelinks]{hyperref}

\title{\textbf{Pose-Aware Multimodal Automatic Tagging\\for Greek Traditional Music}}

\author{
    Alexandros Alexiou\textsuperscript{1}
    \and
    Charilaos Papaioannou\textsuperscript{1,2}
    \and
    Alexandros Potamianos\textsuperscript{1}
    \\[0.6em]
    \small
    \textsuperscript{1}School of Electrical and Computer Engineering,\\
    \small
    National Technical University of Athens, Athens, Greece
    \\[0.4em]
    \small
    \textsuperscript{2}Institute for Language and Speech Processing,\\
    \small
    Athena Research Center, Athens, Greece
    \\[0.4em]
    \small
    \texttt{alexioualexis@hotmail.com}
}

\date{}

\begin{document}

\maketitle

\begin{abstract}
Automatic tagging is a core task in Music Information Retrieval (MIR), yet most tagging systems exploit only audio. Live music performance is inherently multimodal, as semantic labels such as instruments, regional styles, and dance forms are encoded simultaneously across acoustic, visual, and embodied performance cues. This is especially true of culturally specific repertoires such as Greek traditional music, which remain underrepresented in MIR benchmarks. In this paper, we investigate whether the use of dancer pose provides complementary information for automatic tagging in Greek traditional music beyond audio. Using the Lyra dataset, we extend prior audio-only work by extracting aligned video features and pose-derived skeleton streams, enabling an experimental setting for multimodal auto-tagging. We further introduce an automated pipeline for extracting primary-dancer skeleton sequences from in-the-wild dance footage, combining dance-scene detection, multi-person tracking, dancer selection, pose estimation, and quality filtering. We compare unimodal, all bimodal combinations, and trimodal systems using multiple fusion strategies. Audio remains the strongest single modality (AST: macro ROC-AUC 0.821), while skeletons, though weak in isolation, enhance performance through multimodal fusion. The best trimodal system improves macro ROC-AUC by about 4 percentage points over the strongest audio baseline.
\end{abstract}

\section{Introduction}\label{sec:introduction}

Most tagging systems operate on audio alone, even though live music is
an inherently embodied practice in which performers move, audiences
dance, and instruments are visible alongside sound. Consequently,
audio, while a strong baseline, may not always capture all
performance-related cues.
This is particularly important for living folk traditions, where 
cultural identity is articulated through movement, appearance, and instrumentation besides sound; a
regional dance style, a specific instrument, or a characteristic
costume may all encode the same semantic information as tags of a specific genre or geographic origin.

In this setting, the movement of dancers can be embodied by human poses.
A \textit{pose} describes a body configuration at a particular time instance, while a \textit{pose stream} defines a temporally ordered sequence of poses. 
Poses are typically represented as skeleton keypoints, defined in the 2D or 3D domain. 
Skeleton data offer a lightweight, computationally efficient, and privacy-preserving representation, which capture movement patterns when obtained over time, enabling spatio-temporal modeling of human motion.
Simultaneously, they neglect visual cues that could be important (e.g., appearance, objects) in live music performance environments.
This raises a concrete question:
\emph{can poses of dancers  
provide useful complementary information
beyond audio for automatic tagging in culturally specific music?}

In this paper, we address this question by introducing a pose-aware
multimodal approach for automatic tagging in Greek traditional music.
For this purpose, we use the Lyra dataset~\cite{lyra2022}, which indexes
1570 videos from the Greek television program \textit{To Alati tis Gis}
and is annotated with multilabel tags for instruments, genres, and
geographic origins. Prior work on Lyra~\cite{westtoeast2023} established
strong audio-based baselines and explored cross-cultural transfer in an
audio-only setting. In contrast, this work investigates whether the
visual and embodied dimensions of these recordings can provide
complementary information for culturally specific music tagging.

We make three contributions.
\textbf{(1) A pose-aware multimodal tagging setting for Lyra.}
We extend the prior audio-only use of Lyra by extracting aligned video
representations and dancer pose streams, 
enabling the first
systematic study of audio, video, and skeleton data for Greek
traditional music automatic tagging.
\textbf{(2) An automated pose extraction pipeline for in-the-wild
dance footage.}
We introduce a pipeline combining dance-scene detection, multi-person
tracking, primary-dancer selection, pose estimation, and quality
filtering, designed for unconstrained broadcast recordings rather than
curated motion data.
\textbf{(3) A category-level evaluation of multimodal tagging performance.}
We provide a category-level analysis of multimodal tagging performance
across \textit{instruments}, \textit{genres}, and \textit{geographic origins}, allowing us to
examine how audio, video, and skeleton data representations contribute
to different types of semantic music tags.
Code and pre-extracted features are being made available\footnote{\href{https://github.com/alexioualex9/Pose-Aware-Multimodal-Automatic-Tagging}{\texttt{https://github.com/alexioualex9/Pose-Aware-Multimodal-Automatic-Tagging}}} to support reproducibility of all reported experiments.

\section{Related Work}\label{sec:related}

\textbf{Music tagging in world music datasets.}
Automatic music tagging has been studied extensively with audio-only
models on large-scale datasets~\cite{vgg_ish,gong2021ast,abu2016youtube8m}.
At the same time, widely used MIR benchmarks remain centered on Western
repertoires~\cite{bertin2011msd,goto2002rwc}, while culturally specific
and non-Western traditions are comparatively underrepresented~\cite{serra2011multicultural}.
Lyra~\cite{lyra2022} is one of the few resources designed for Greek
traditional music, and prior work has so far focused on audio-based
representation learning and cross-cultural transfer~\cite{westtoeast2023}.

\textbf{Multimodal music information retrieval.}
Prior multimodal MIR work has explored genre classification,
cross-modal retrieval, and audio--video fusion for music performance
understanding~\cite{oramas2018multimodal,li2019query,christodoulou2025musiqal}.
These studies suggest that visual information can complement audio in
performance-centered settings, but they have not focused on multilabel
autotagging in culturally specific music collections.

\textbf{Pose-based modeling for music and dance.}
Pose-based sequence modeling has been shaped by ST-GCN and its
extensions~\cite{yan2018stgcn,shi2019twostream,liu2020disentangling}. In music and dance applications, pose representations have been used for
dance generation~\cite{siyao2022bailando}, beat tracking~\cite{pedersoli2020dance},
and performance analysis~\cite{godoy2009body,lerch2020mpa,clayton2022raga,roychowdhury2025multimodal,clayton2024hindustani}.
However, skeleton data have not been systematically explored for
multilabel music autotagging, especially in unconstrained broadcast footage.

\section{Dataset and Pose Extraction}\label{sec:data}

\subsection{The Lyra Dataset}

Lyra provides annotations for Greek traditional music videos recorded
from multiple camera angles at 25 fps, with a frame resolution of
$690{\times}380$ px. The content spans dance scenes, musicians,
singers, and audience shots. Each video is annotated with the non-exclusive tags: \textit{instrument}, \textit{genre}, and \textit{geographic-origin tags}.

Lyra comprises 1570 audio-video tracks, of which 767 contain dance content.
The pose extraction pipeline described in Section~\ref{ssec:skeleton}
yields 749 tracks with valid pose streams, forming the
\textbf{dance subset} used in this study.
The predefined train/validation/test split provided with the Lyra dataset
repository is inherited unchanged and restricted to this subset, yielding
a 533/62/154 split.
Following prior work~\cite{westtoeast2023}, which evaluated the top-30 most frequent labels,
we begin from the same label set; however, two of these labels have no
positive instances in the dance subset, so evaluation is conducted on
the remaining 28 labels.

\subsection{Pose Extraction Pipeline}
\label{ssec:skeleton}

Our pipeline detects dance scenes, extracts primary-dancer poses within them,
and converts the resulting pose sequences into quality-filtered skeleton
tensors.

\begin{figure}[t]
  \centering
  \begin{minipage}[b]{0.5\linewidth}
    \centering
    \includegraphics[width=\linewidth]{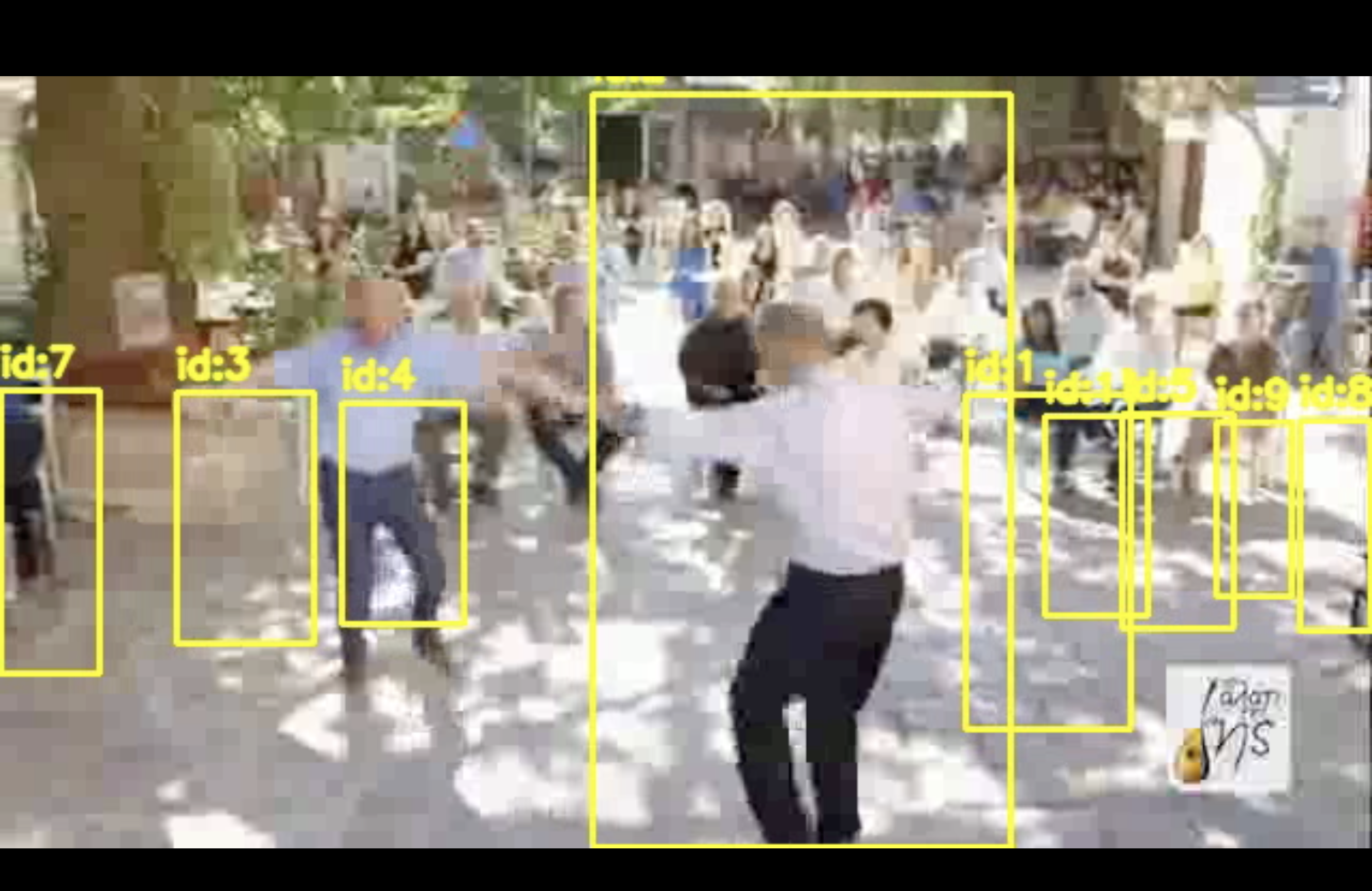}
    \small (a) ByteTrack multi-person tracking
  \end{minipage}%
  \hfill
  \begin{minipage}[b]{0.35\linewidth}
    \centering
    \includegraphics[width=\linewidth]{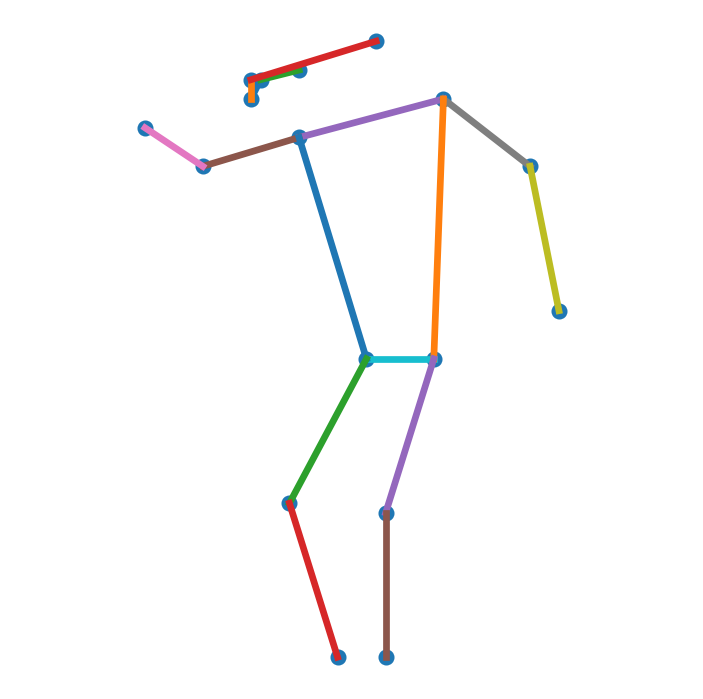}
    \small (b) Extracted skeleton graph
  \end{minipage}
  \caption{(a) ByteTrack assigns IDs to all detected  
               persons; the primary dancer is selected 
               by maximising bounding-box area times
               detection confidence
           (b) AlphaPose yields a 17-joint COCO skeleton; colour-coded
           bones form the spatial graph for ST-GCN.}
  \label{fig:skeleton_pipeline}
\vspace{-0.2cm}
\end{figure}

\textbf{Dance scene detection.} We first identify dance scenes within Lyra. To train the dance-scene
detector, we randomly sampled 10 videos from the Lyra training split and
manually annotated 300 one-second video segments in total, balanced between dance and
non-dance content. The segments were split into train/validation/test sets,
ensuring no video overlap across splits.
Fine-tuning only the final classification layer of an MViT model~\cite{fan2021mvit} 
pretrained on Kinetics-400~\cite{kay2017kinetics} yielded 98\% segment-level accuracy on the held-out test set. At inference time, scene-level labels were obtained
by majority voting over one-second predictions within each
PySceneDetect~\cite{castellano2025pyscenedetect} segment and used to guide pose extraction.

\textbf{Pose extraction.}
Within detected dance scenes, ByteTrack~\cite{zhang2022bytetrack} with a YOLO
detector~\cite{redmon2016yolo} tracks all individuals across frames.
At each frame, we identify the primary dancer using an area-based
person-selection heuristic inspired by prior work on person selection
for action recognition~\cite{jacquot2020subtle}. Specifically, we select
$j^* = \arg\max_j (\operatorname{area}(b_j) \cdot c_j)$,
where $b_j$ denotes the bounding box of person $j$ and $c_j$ its
associated detection confidence score.
We restrict pose
extraction to a single primary dancer as the broadcast footage is highly
unconstrained, often containing audience members, musicians, singers, and
partially visible persons, which makes multi-person pose modeling
substantially more error-prone. The selected ByteTrack bounding boxes are then used to crop the primary dancer
and pass the crops to AlphaPose~\cite{fang2023alphapose}, which outputs
$V=17$ COCO~\cite{lin2014coco} body keypoints $(x,y,c)$ per frame, with 2D image coordinates and
confidence scores.

\textbf{Pose stream representation.}
We convert extracted dancer poses into pose stream representations over
fixed-length clips. After hip-centering, we reduce pose-estimation outliers
with a self-similarity filter removing poses deviating from the clip-level
medoid~\cite{kaufman1987clustering}. The remaining poses are assigned a
pose-quality score combining joint confidence, bone-length consistency,
left--right symmetry, and temporal jitter
penalties~\cite{vemulapalli2014human,hoang2024robust,dabral2018structure,cao2020anatomy,hossain2018exploiting}.
Based on this score, we retain up to $N_p=32$ valid dancer poses per clip
and preserve their temporal order. Clips with fewer than 8 valid poses are
discarded, while clips with 8 to 31 are padded to $N_p=32$.
For each retained pose, we stack position $(x,y,c)$ and first-order motion
$(\Delta x,\Delta y,c)$ features~\cite{yan2018stgcn,shi2019twostream}, preserving confidence
$c$ in both streams for joint reliability, and obtain $V$-joint, $C$-channel tensors
\begin{equation}
  X \in \mathbb{R}^{C \times N_p \times V}, \; C{=}6,\; N_p{=}32,\; V{=}17,
\label{eq:skeleton_tensor}
\end{equation}
where $C$ is the number of channels.

\section{Methodology}\label{sec:methodology}

\subsection{Temporal Segmentation and Modality Alignment}


\begin{figure}[t]
  \centering
  \includegraphics[width=1.00\linewidth]{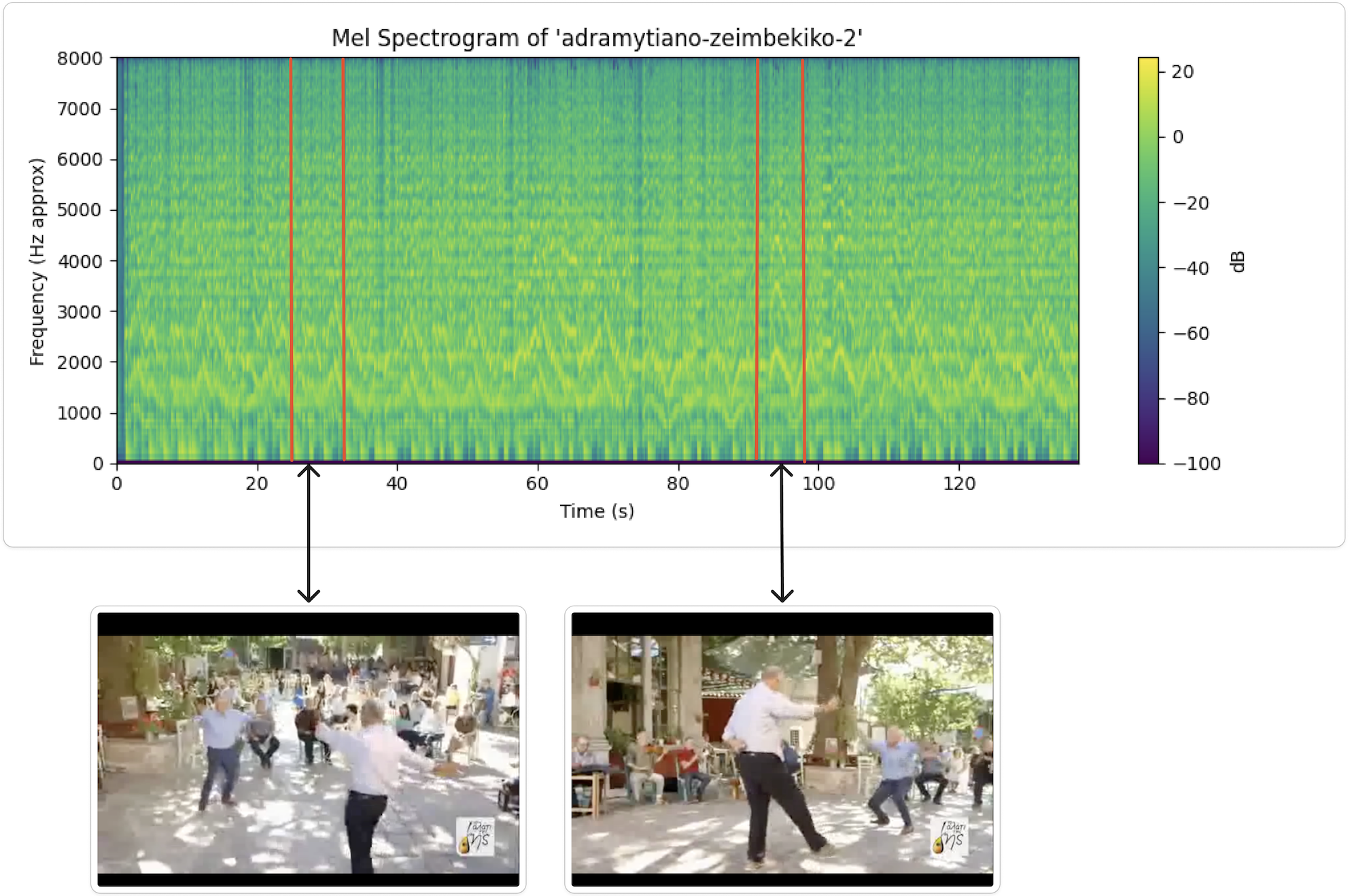}
  \caption{Temporal alignment of audio, video, and skeleton clips for
           \textit{adramytiano-zeimbekiko-2}. Red lines on the mel-spectrogram
           mark clip boundaries, and arrows indicate the corresponding sampled
            video frames.}
  \label{fig:alignment}
\vspace{-0.2cm}
\end{figure}

All modalities are segmented into shared non-overlapping clips of
duration $T_{\mathrm{clip}}{=}8.00$~s, which define the alignment units for
multimodal fusion. For audio, each clip is represented by a
mel-spectrogram computed at $f_s{=}16\,000$~Hz with hop size $h{=}256$
samples. This yields
$N_f=\operatorname{round}(T_{\mathrm{clip}} f_s/h)=500$ time steps per clip,
with duration $\delta_f=h/f_s=16$~ms. The $k$-th clip therefore spans
timestamps
\begin{equation}
t_\text{start}^{(k)} = k \cdot N_f \cdot \delta_f, \qquad
t_\text{end}^{(k)} = (k+1) \cdot N_f \cdot \delta_f .
\end{equation}
The same temporal boundaries are used for video and skeleton feature
extraction, aligning audio, video, and skeleton data.
Figure~\ref{fig:alignment} illustrates
the alignment procedure: red vertical lines on the mel-spectrogram denote
example clip boundaries, and one
representative video frame is shown for each clip.

\subsection{Unimodal Representation Learning}
\label{unimodals}

\begin{figure*}[t]
  \centering
  \includegraphics[width=0.95\linewidth]{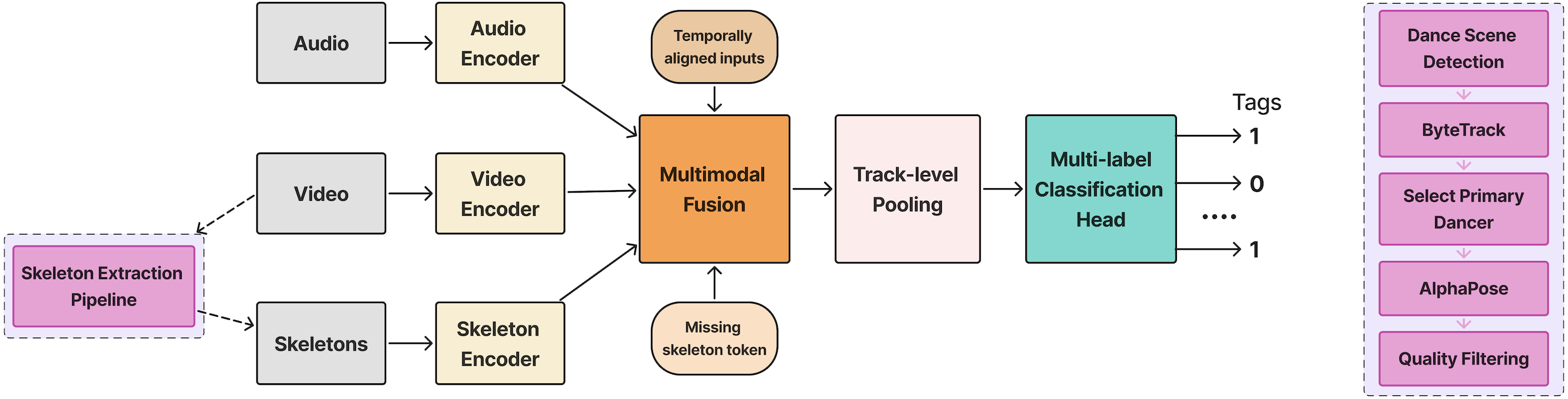}
  \caption{Overview of the proposed multimodal tagging pipeline. Solid arrows denote
  the main tagging flow, in which temporally aligned audio, video, and
  skeleton clips are encoded, fused, and pooled for automatic tagging. Dashed arrows indicate the pose-extraction
  path and the expanded pink block on the right
  details the full pipeline as described in \ref{ssec:skeleton}.
  }
  \label{fig:architecture}
\vspace{-0.2cm}
\end{figure*}

We model each modality with a dedicated encoder operating on temporally
aligned clips (Figure~\ref{fig:architecture}). For each clip, the audio, video,
and, when available, skeleton encoders produce clip-level representations.
We next describe the unimodal encoder for each modality.

\textbf{Audio}. We use the Audio Spectrogram Transformer
(AST)~\cite{gong2021ast} as an audio encoder; this model has been shown to serve as a strong baseline for
Lyra~\cite{westtoeast2023}. AST treats the input mel-spectrogram as a sequence of patches and applies
ViT-style self-attention over 8-second windows, allowing the model to
capture long-range tonal and rhythmic dependencies. Each audio clip is mapped to an embedding
$\mathbf{z}_a \in \mathbb{R}^{D_a}$, where $D_a$ denotes the audio
embedding dimension. We additionally experimented with
VGGish~\cite{vgg_ish}, using 3.69-second clip windows following prior
work on Lyra~\cite{westtoeast2023}, but it did not outperform AST and is therefore not
considered further.

\textbf{Video}. We consider two families of visual encoders, namely \textit{image-based}
and \textit{video-based}. For the \textit{image-based} setting, ResNet-50~\cite{he2016resnet}
and ViT-B/16~\cite{dosovitskiy2021vit} were selected. These models encode sampled RGB frames
independently with a visual encoder
$f_\theta:\mathbb{R}^{H\times W\times 3}\rightarrow\mathbb{R}^{D_v}$, and obtain a
clip-level representation by mean pooling the frame embeddings:
\begin{equation}
\mathbf{z}_v = \frac{1}{N}\sum_{n=1}^{N} f_\theta(\mathbf{x}_n),
\end{equation}
where $\mathbf{x}_n \in \mathbb{R}^{H\times W\times 3}$ denotes the $n$-th sampled
RGB frame, $N$ is the number of sampled frames in the clip, $\theta$ denotes the
encoder parameters, and $\mathbf{z}_v\in\mathbb{R}^{D_v}$ is the resulting
clip-level visual embedding of dimension $D_v$.
For the \textit{video-based} setting, SlowFast-50~\cite{feichtenhofer2019slowfast},
R(2+1)D~\cite{tran2018r21d}, and TimeSformer~\cite{bertasius2021timesformer} were employed.
These architectures operate on video segments and
explicitly model temporal dynamics, producing clip-level visual representations
directly. All visual backbones are initialized from standard ImageNet~\cite{russakovsky2015ilsvrc}
or Kinetics pretrained checkpoints and are followed by a classification head.

\textbf{Skeletons}. We use a lightweight ST-GCN-style model~\cite{yan2018stgcn} as the skeleton encoder. ST-GCN provides an established formulation
for skeleton-based modeling, making it a suitable architectural basis for
assessing the contribution of pose in our setting. We therefore adopt a
variant tailored to our dancer poses. The 17 body joints are represented
as graph nodes, with anatomical connections defining the graph edges.
Following the standard ST-GCN partitioning scheme, we use a fixed
3-subset adjacency for self, centripetal, and centrifugal connections.
The network comprises 6 graph-convolutional blocks with 64 channels,
multi-scale temporal convolutions with kernel sizes 9 and 3, residual
connections, and global average pooling followed by a linear classification
head. The global pooling stage produces a clip-level skeleton embedding
$\mathbf{z}_s \in \mathbb{R}^{D_s}$, where $D_s$ denotes the skeleton
embedding dimension. In contrast to deeper adaptive variants, the
adjacency remains fixed throughout, which keeps the model lightweight
and reduces the risk of overfitting on the relatively small and noisy
dance subset. In the main experiments, the skeleton encoder is trained on 3.69-second
windows, following the shorter window duration used for VGGish in prior
Lyra experiments~\cite{westtoeast2023}. As shown in the ablation study in
Section~\ref{subsec:skeleton_ablation_study}, this shorter duration yields more robust pose
representations than training directly on 8.00-second windows. For
multimodal fusion, the trained encoder is applied to the pose stream
associated with each aligned 8.00-second clip, using the tensor
representation defined in \eqref{eq:skeleton_tensor}, producing one
skeleton embedding per clip.

\subsection{Multimodal Fusion}


Given the clip-level unimodal representations defined in Section~\ref{unimodals}, we fuse the
available modalities using several multimodal fusion strategies. We evaluate
late fusion, which averages track-level probabilities from trained unimodal
baselines, and learned early-fusion, gated-fusion, and cross-attention models
operating on fixed pre-classifier embeddings from frozen modality-specific
encoders. The learned architectures share the same training setup but differ in where
cross-modal interaction occurs; performance differences therefore reflect
the full architecture rather than the fusion operator alone. For skeleton-inclusive models, inspired by prior work on
missing-modality handling in multimodal learning~\cite{ramazanova2024mmt}, we use an availability mask and
preserve temporal alignment in clips without valid dancer poses by using a
learnable missing-skeleton token in place of the transformed pose
representation. We describe each fusion strategy in turn.

\textbf{Late fusion.}
As a simple decision-level fusion baseline, we combine modality-specific
track-level probabilities using late fusion~\cite{atrey2010multimodal,baltrusaitis2019multimodal}.
For each modality $m$, clip-level outputs are first temporally aggregated
to obtain a track-level prediction $\hat{\mathbf{p}}^{(m)}$.
For audio and video, aggregation uses the aligned clip sequence, whereas
for skeleton it uses only clips with valid pose estimates. The final
multimodal prediction is obtained as
\begin{equation}
\hat{\mathbf{p}} = \sum_{m} w_m \hat{\mathbf{p}}^{(m)}.
\end{equation}
We use uniform weights, setting $w_m = 1/2$ for bimodal fusion and
$w_m = 1/3$ for trimodal fusion, yielding a parameter-free fusion rule without validation-set weight tuning.

\textbf{Early fusion.}
Modality embeddings are projected to a shared latent space and arranged
as tokens, with one token per modality per timestep and missing skeletons
represented by the learnable missing-skeleton token. The resulting token
sequence is processed by a Transformer encoder~\cite{vaswani2017attention}, with a
learnable $\text{CLS}$ token used for classification. Inspired by
multimodal Transformers such as VisualBERT~\cite{li2019visualbert}, this
formulation enables joint self-attention across both time and modality.

\textbf{Gated fusion (GMU).}
We also consider a
GMU-style fusion module~\cite{arevalo2017gmu}. For each clip, each modality is
projected to a hidden representation, and learned gates
determine its contribution to the fused token. In the bimodal case, a
single gate weights the two modality-specific hidden states,
whereas in the trimodal case we use one sigmoid gate per modality. When
skeleton is missing, its branch is replaced by a learned
missing-skeleton token, whose hidden state is treated as a regular modality input by the gating mechanism. The resulting fused
sequence is passed to a Transformer encoder for temporal modeling and
track-level prediction.

\textbf{Cross-attention.}
A MulT-style cross-attention model is studied for directional
cross-modal interaction~\cite{tsai2019multimodal}. Modality embeddings
are projected to a shared latent space, and each target stream is
updated via pairwise cross-attention from the others. The resulting
sequences are processed by temporal encoders, and the final temporal states
are concatenated for multilabel prediction. In skeleton-inclusive
settings, a learnable missing-skeleton token is used when skeleton is a
target stream, while skeleton serves as a source stream only at valid-pose timesteps.

\section{Experiments}\label{sec:experiments}

All results are reported as mean\,$\pm$\,std over 5 random seeds.
Evaluation is performed at the track level throughout: clip-level logits
are mean-pooled per track, sigmoid is applied to obtain track-level label
probabilities, and all metrics are macro-averaged over tags, giving equal
weight to each tag. We report ROC-AUC, which measures the ranking quality
of positive versus negative examples; PR-AUC, which summarizes the
precision--recall trade-off~\cite{davis2006pr}; and F1, the harmonic mean of
precision and recall. F1 is computed from binarized track-level predictions
using a fixed threshold of 0.5 for all labels and models.

For the \textbf{audio} unimodal experiments, we use pretrained input
normalization and train with BCEWithLogitsLoss, Adam~\cite{kingma2015adam},
a batch size of 12, a learning rate of $10^{-5}$, and a MultiStep scheduler
for up to 50 epochs, with early stopping (patience of 5 validation epochs).
During training, clips are sampled randomly; at test time, predictions are
aggregated over all non-overlapping clips.

For the \textbf{video} unimodal experiments, we use pretrained backbones with a shared
two-layer MLP head (hidden size 1024, GELU~\cite{hendrycks2016gelu}, dropout), and are
trained with BCEWithLogitsLoss, AdamW~\cite{loshchilov2019adamw}, a 
batch size of 512, a learning rate of $10^{-3}$, and a cosine decay 
scheduler for up to 200 epochs, with early stopping (patience of 8 validation epochs). During training, clips are sampled randomly, 
whereas validation and test use all available clips. Backbones remain 
frozen, and only the classification head is trained. Partial fine-tuning of 
the last backbone block did not consistently improve validation 
performance; therefore, we report only frozen-backbone results.

For the \textbf{skeleton} unimodal experiments, the skeleton model was trained with BCEWithLogitsLoss,
AdamW, a batch size of 32, a learning
rate of $3 \times 10^{-4}$, and 30 epochs without early stopping.
Because valid pose observations are available only for a subset of clips,
training and evaluation use all clips with valid dancer poses, and
track-level predictions are obtained by pooling over the corresponding
clip-level outputs.

For all \textbf{fusion} experiments, we use fixed SlowFast-50 video embeddings and train the fusion models with AdamW, BCEWithLogitsLoss, a batch 
size of 64 and a learning rate of $2 \times 10^{-4}$ for 60 epochs without 
early stopping. We use a shared latent dimension of $D=256$, one 
Transformer layer and 4 attention heads.

\section{Results}\label{sec:results}

\begin{table}[t]
\centering
\caption{Video model results on the dance subset (top-28 labels), with
macro metrics reported as mean\,$\pm$\,std over 5 seeds.
\textbf{Bold} marks the best per column.}
\label{tab:video_full}
\small
\setlength{\tabcolsep}{2pt}
\begin{tabular*}{\columnwidth}{@{\extracolsep{\fill}}lccc@{}}
\toprule
\cmidrule(lr){2-4}
\textbf{Model} & \textbf{ROC-AUC} & \textbf{PR-AUC} & \textbf{F1} \\
\midrule
SlowFast-50
  & $\mathbf{0.759_{\pm 0.020}}$ & $\mathbf{0.435_{\pm 0.006}}$ & $0.238_{\pm 0.015}$ \\
TimeSformer
  & $0.734_{\pm 0.023}$ & $0.406_{\pm 0.025}$ & $\mathbf{0.251_{\pm 0.020}}$ \\
R(2+1)D
  & $0.685_{\pm 0.012}$ & $0.341_{\pm 0.010}$ & $0.195_{\pm 0.017}$ \\
ResNet-50
  & $0.666_{\pm 0.004}$ & $0.318_{\pm 0.003}$ & $0.202_{\pm 0.006}$ \\
ViT-B/16
  & $0.677_{\pm 0.007}$ & $0.325_{\pm 0.014}$ & $0.185_{\pm 0.004}$ \\
\bottomrule
\end{tabular*}
\end{table}

\begin{table}[t]
\centering
\caption{Full multimodal results on the dance subset (top-28 labels),
with macro metrics reported as mean\,$\pm$\,std over 5 seeds.
The top block reports unimodal baselines (A\,=\,audio, V\,=\,best video model,
S\,=\,skeleton) once for reference.
\textbf{Bold} marks the best per column within each fusion block
(fusion rows only).}
\label{tab:trimodal_full}
\small
\resizebox{0.65\textwidth}{!}{%
\renewcommand{\arraystretch}{1.2}
\begin{tabular}{llccc}
\toprule
\cmidrule(lr){3-5}
\textbf{Fusion} & \textbf{Method} & \textbf{ROC-AUC} & \textbf{PR-AUC} & \textbf{F1} \\
\midrule
\textit{A}   & \textit{AST} & $0.821_{\pm 0.019}$ & $0.477_{\pm 0.027}$ & $0.328_{\pm 0.021}$ \\
\textit{V}   & \textit{SlowFast-50}  & $0.759_{\pm 0.020}$ & $0.435_{\pm 0.006}$ & $0.238_{\pm 0.015}$ \\
\textit{S}   & \textit{ST-GCN}  & $0.586_{\pm 0.008}$ & $0.287_{\pm 0.007}$ & $0.193_{\pm 0.007}$ \\
\midrule
A+V   & Late       & $\mathbf{0.852_{\pm 0.002}}$ & $\mathbf{0.508_{\pm 0.005}}$ & $0.263_{\pm 0.003}$ \\
      & Early      & $0.843_{\pm 0.005}$          & $0.504_{\pm 0.010}$          & $0.394_{\pm 0.035}$ \\
      & Gated      & $0.849_{\pm 0.006}$          & $0.498_{\pm 0.008}$          & $0.401_{\pm 0.026}$ \\
      & Cross-Attn & $0.850_{\pm 0.014}$          & $0.495_{\pm 0.027}$          & $\mathbf{0.410_{\pm 0.025}}$ \\
\midrule
A+S   & Late       & $0.811_{\pm 0.004}$          & $0.442_{\pm 0.008}$          & $0.253_{\pm 0.008}$ \\
      & Early      & $0.832_{\pm 0.011}$          & $0.487_{\pm 0.018}$          & $0.367_{\pm 0.038}$ \\
      & Gated      & $\mathbf{0.839_{\pm 0.003}}$ & $\mathbf{0.498_{\pm 0.006}}$ & $\mathbf{0.411_{\pm 0.026}}$ \\
      & Cross-Attn & $0.831_{\pm 0.007}$          & $0.477_{\pm 0.024}$          & $0.400_{\pm 0.029}$ \\
\midrule
V+S   & Late       & $0.716_{\pm 0.007}$          & $0.381_{\pm 0.014}$          & $0.206_{\pm 0.012}$ \\
      & Early      & $0.781_{\pm 0.009}$          & $0.501_{\pm 0.021}$          & $0.396_{\pm 0.020}$ \\
      & Gated      & $\mathbf{0.798_{\pm 0.007}}$ & $\mathbf{0.505_{\pm 0.018}}$ & $\mathbf{0.405_{\pm 0.026}}$ \\
      & Cross-Attn & $0.797_{\pm 0.008}$          & $0.455_{\pm 0.022}$          & $0.366_{\pm 0.024}$ \\
\midrule
A+V+S & Late       & $0.842_{\pm 0.002}$          & $0.490_{\pm 0.012}$          & $0.234_{\pm 0.015}$ \\
      & Early      & $0.847_{\pm 0.008}$          & $0.497_{\pm 0.006}$          & $0.401_{\pm 0.022}$ \\
      & Gated      & $\mathbf{0.864_{\pm 0.003}}$ & $\mathbf{0.525_{\pm 0.012}}$ & $0.393_{\pm 0.019}$ \\
      & Cross-Attn & $0.853_{\pm 0.009}$          & $0.514_{\pm 0.028}$          & $\mathbf{0.422_{\pm 0.027}}$ \\
\bottomrule
\end{tabular}}
\end{table}

Throughout this section, we use A, V, and S to denote the audio, video,
and skeleton modalities, respectively; multimodal combinations follow the
same notation, e.g., A+V, A+S, V+S, and A+V+S.

\subsection{Unimodal Baselines}

Among the video models, SlowFast-50 performs best on ranking metrics,
achieving the highest macro ROC-AUC and PR-AUC, while TimeSformer remains
competitive, as shown in Table~\ref{tab:video_full}. Based on our results, audio has the strongest unimodal signal overall. Audio baseline achieves
macro ROC-AUC\,0.821, clearly above V (baseline) model (0.759), and also leads on macro PR-AUC and macro-F1
in Table~\ref{tab:trimodal_full}. This suggests that Lyra tags are encoded
primarily in the acoustic channel, while the visual stream provides
complementary evidence.

In contrast, the S (baseline) model performs markedly worse than
the A and V baselines, reaching a macro ROC-AUC of 0.586. This
indicates that pose alone is not sufficient for reliable autotagging;
however, its non-trivial performance suggests that dancer poses still
capture useful cues that can become beneficial under multimodal fusion.

\subsection{Multimodal Fusion Results}

Based on the unimodal video results, we use SlowFast-50 as the video
backbone in all fusion experiments, fixing the visual encoder across
multimodal settings. Under this setup, learned multimodal fusion
improves over the unimodal baselines, as reported in Table~\ref{tab:trimodal_full}.
Relative to the strongest unimodal baseline (A, 0.821 macro ROC-AUC), the best A+V+S reaches 0.864 macro ROC-AUC (+4.3).
More importantly, adding skeletons to the best A+V configuration improves macro ROC-AUC from 0.852 to 0.864 (+1.2).
A similar pattern is observed for A+S and V+S models.
The best A+S model achieves 0.839, outperforming the A (unimodal) baseline
(0.821), while adding skeletons to the V (unimodal) baseline improves performance from 0.759 to 0.798.

A clear pattern also emerges across fusion strategies. Across
modality combinations, Gated fusion is generally strongest on macro
ROC-AUC and PR-AUC, whereas Cross-Attention often yields the best
macro-F1. Late fusion
remains competitive on threshold-independent metrics as it gives the best
A+V macro ROC-AUC and PR-AUC, but is consistently weaker on macro-F1,
suggesting that learned fusion is useful for modeling cross-modal
interactions and improving thresholded predictions.

At the label level, we compare the S baseline against the A and V baselines
to identify cases where skeleton cues are informative, especially for labels
associated with movement or performance practice.
Although weaker than audio overall, the S (unimodal) model achieves higher
F1 on the \textit{Aegean} genre label (0.21 vs. 0.17), possibly
reflecting regionally distinctive dance movement~\cite{lykesas2016}
that is less directly captured by audio.
It also remains competitive on \textit{Voice}
instrument (0.75 vs. 0.77) and \textit{Percussion} instrument (0.82
vs. 0.92). Compared to
video, S reaches
similar performance on labels such as
\textit{Traditional} genre (0.98 vs. 0.98) and \textit{Laouto}
instrument (0.66 vs. 0.70), while remaining weaker on labels such as
\textit{Epirus} place and \textit{Klarino} instrument. These patterns suggest that skeleton does not replace audio or video, but
provides complementary evidence for labels where movement patterns carry
semantic information not fully captured by acoustic or visual appearance
cues alone.

\subsection{Skeleton Robustness and Missingness Ablations}
\label{subsec:skeleton_ablation_study}

\begin{table}[t]
\centering
\caption{Skeleton-specific ablations. Top: sensitivity of the unimodal
skeleton encoder to temporal window duration. Bottom: effect of
missing-skeleton handling in the trimodal gated fusion model. Macro
metrics are reported as mean\,$\pm$\,std over 5 seeds.}
\label{tab:skeleton_ablations}
\small
\setlength{\tabcolsep}{2pt}
\begin{tabular*}{\columnwidth}{@{\extracolsep{\fill}}lccc@{}}
\toprule
\textbf{Setting} & \textbf{ROC-AUC} & \textbf{PR-AUC} & \textbf{F1} \\
\midrule
\multicolumn{4}{l}{\textit{Windowing (skeleton-only)}} \\
Tr 3.69s / Te 3.69s & $\mathbf{0.596_{\pm 0.008}}$ & $\mathbf{0.289_{\pm 0.011}}$ & $\mathbf{0.206_{\pm 0.005}}$ \\
Tr 8.00s / Te 8.00s & $0.555_{\pm 0.010}$ & $0.275_{\pm 0.011}$ & $0.192 _{\pm 0.006}$ \\
Tr 3.69s / Te 8.00s$^\star$ & $0.586_{\pm 0.008}$ & $0.287_{\pm 0.007}$ & $0.193_{\pm 0.007}$ \\
\midrule
\multicolumn{4}{l}{\textit{Missing-skeleton handling (A+V+S Gated)}} \\
Zero-fill     & $0.860_{\pm 0.005}$ & $0.515_{\pm 0.015}$ & $0.384_{\pm 0.032}$ \\
Learned token & $\mathbf{0.864_{\pm 0.003}}$ & $\mathbf{0.525_{\pm 0.012}}$ & $\mathbf{0.393_{\pm 0.019}}$ \\
\bottomrule
\end{tabular*}
\end{table}

We further examine two skeleton-specific design choices. First, we assess
the effect of temporal window duration for the unimodal skeleton encoder.
Training and evaluating on 3.69-second windows yields the strongest
skeleton-only macro ROC-AUC (0.596), whereas using 8.00-second windows
throughout reduces performance to 0.556, as shown in Table~\ref{tab:skeleton_ablations}. This suggests that longer
windows introduce additional noise for skeleton-based tagging. When the
encoder trained on 3.69-second windows is applied to pose streams from
the aligned 8.00-second clips, performance remains close to the
shorter-window setting (0.586) and clearly exceeds direct 8.00-second
training. This supports our use of a 3.69-second skeleton encoder within
the aligned 8.00-second multimodal pipeline.

Second, we ablate missing-skeleton handling in the best trimodal gated
fusion model. Replacing missing skeleton clips with zeros yields
slightly weaker results (ROC-AUC 0.860, PR-AUC 0.515, F1 0.384) than
the learned missing-skeleton token (0.864 / 0.525 / 0.393), as reported in Table~\ref{tab:skeleton_ablations}, indicating
a modest but consistent benefit over naive zero substitution.

\subsection{Category-Level Analysis}

\begin{figure}[h]
  \centering
  \includegraphics[width=0.75\linewidth]{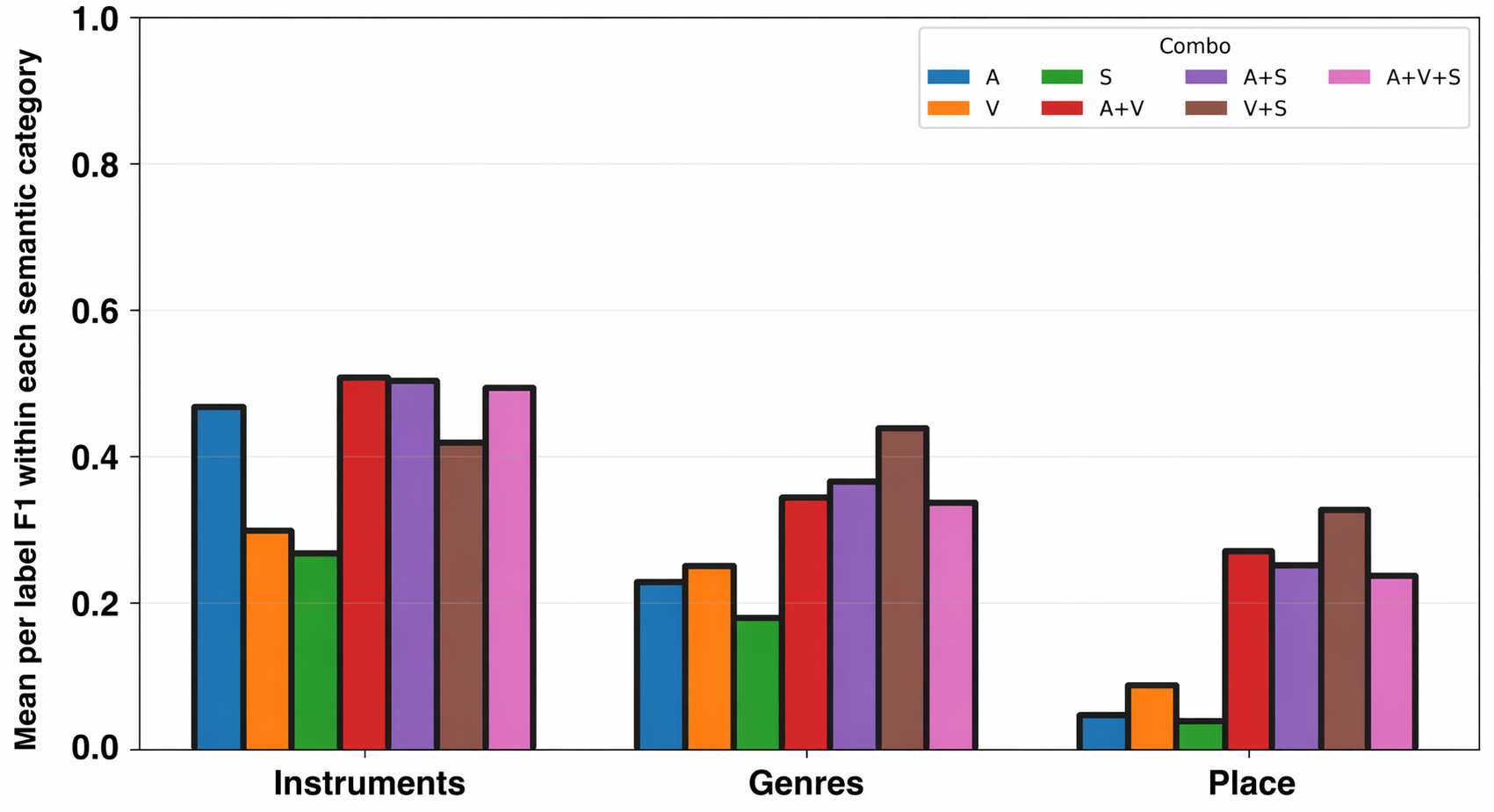}
  \caption{Mean F1 by tag category (instruments, genre, and geographic-origin) for all unimodal models and the best model from each multimodal combination.}
  \label{fig:type_tag_results}
\vspace{-0.2cm}
\end{figure}

For category-level analysis, we compute the mean per-label F1 within
each semantic category:
\begin{equation}
\mathrm{F1}_{g,r}
=
\frac{1}{|L_g|}
\sum_{\ell \in L_g}
\mathrm{F1}_{\ell,r},
\label{eq:category_f1}
\end{equation}
where $L_g$ is the set of labels in category $g$ and
$\mathrm{F1}_{\ell,r}$ is the F1 score of label $\ell$ for modality
setting $r$.

The three Lyra tag categories (instruments, musical genres, and
geographic-origin) differ substantially in how well they are captured
by each modality. Instrument tags are primarily grounded in acoustics
and visible instrument appearance, whereas genre and geographic-origin
tags may also reflect embodied movement and performance practice. We
therefore expect motion cues to be especially relevant for genre and geographic-origin labels.

This tendency is reflected in the category-level results in
Figure~\ref{fig:type_tag_results}. Instrument tags remain primarily
audio-led, with A+V yielding the highest mean per-category F1 (0.508).
By contrast, genre tags show the strongest relative gains from
visual-motion information, with V+S emerging as the strongest
combination (0.438). 

Geographic-origin tags remain the hardest category overall, but show the
largest relative multimodal gain: V+S reaches 0.324 mean F1, compared with
unimodal scores below 0.086. This suggests that geographic-origin labels,
although difficult to infer from any single modality, benefit substantially
from the combination of visual appearance and skeleton cues.

\section{Conclusion}\label{sec:conclusion}

In this paper, we presented a pose-aware multimodal extension of Lyra for
automatic tagging in Greek traditional music. 
We introduced a pipeline that extracts pose streams from broadcast footage, combines audio, video, and pose stream representations, handles missing pose observations, and evaluates multiple fusion strategies for auto-tagging.

Our results show that audio remains the strongest unimodal signal, while
video and dancer poses contribute complementary information in
multimodal fusion. Although the skeleton model is weaker as a
standalone tagger, pose improves performance when combined with audio or
video. Category-level analysis suggests that
these gains are concentrated on tags for which embodied performance cues are
informative.

These findings are nevertheless shaped by data scale and pose quality: the
dance subset contains 749 videos, and the skeleton pipeline covers about
50\% of clips due to occlusions, multi-person ambiguity, and scene transitions.
Future work could improve multi-person tracking for
ensemble dance scenes, and explore end-to-end multimodal pretraining or
cross-cultural transfer across Lyra's regional subsets to better assess
generalization across dance and music traditions. More broadly, this work demonstrates a viable path toward enriching multimodal cultural archives from broadcast recordings at scale.

\section{AI Usage Statement}

AI-based writing assistance tools were used in the preparation of this manuscript to support text editing and formatting. All scientific content, results, and conclusions are the sole responsibility of the authors.

\section{Ethics Statement}

The Lyra Dataset consists of recordings from the publicly broadcast
Greek television program \textit{To Alati tis Gis} and is used here
for non-commercial research.
The pose extraction pipeline extracts only anonymised 2-D joint coordinates;
no biometric identifiers are stored or released.
Dance and performance footage is used solely to derive motion
features for the music tagging task.

\bibliographystyle{IEEEtran}
\bibliography{ISMIRBibliography}

\end{document}